# Measurements of Spin Polarization of Epitaxial SrRuO$_3$ Thin Films

B. Nadgorny[1], M. S. Osofsky[2], D.J. Singh[2], G.T. Woods[2], R. J. Soulen, Jr.[2], M.K. Lee[3], S.D. Bu[3], C. B. Eom[3].

[1]Dept. of Physics and Astronomy, Wayne State University, Detroit, MI 48201

[2]Naval Research Laboratory, Washington, DC 20375

[3]Dept. of Material Science and Engineering, University of Wisconsin- Madison, WI 53706

Abstract

We have measured the transport spin-polarization of epitaxial thin films of the conductive ferromagnetic oxide, SrRuO$_3$, using Point Contact Andreev Reflection Spectroscopy (PCAR). In spite of the fact that spin-up and spin-down electronic densities of states at the Fermi level for SrRuO$_3$ calculated from band structure theory are practically the same, the experimental transport spin polarization for these films was found to be about 50%. This result is a direct consequence of the Fermi velocity disparity between the majority and minority bands and is in good agreement with our theoretical estimates.

Recently, there has been a renaissance of research in the physics of oxides, and especially perovskite-based materials. While many of these materials are very similar structurally, they exhibit a wide variety of properties including ferromagnetism, superconductivity, ferroelectricity, and metal-insulator transitions. The development of thin film deposition techniques for oxides has helped to exploit their structural similarity and to grow epitaxial multilayer heterostructures comprised of materials with different physical properties. Important advances in fabrication technique allows one now to create novel devices, such as ferroelectric capacitors[1], with $SrRuO_3$ as a metallic layer. In addition to more conventional devices, it has lately become possible to create and study a new class of electronic devices containing ferromagnet/insulator/ferromagnet (FIF) structures, whose properties are controlled by the electron spin of the magnetic material[2,3]. Importantly, $SrRuO_3$ (including its modifications) is the only known ferromagnetic metal (FM) among 4d oxides with magnetization $m \approx 1.6$ $\mu_B$/Ru and Curie temperature $T_C \approx 165$ K. Due to its structural simplicity and remarkable chemical stability, $SrRuO_3$ forms an excellent interface with $Al_2O_3$ barrier layers[4] and thus has attracted interest for studies of spin-dependent transport and as a potential component of these devices. [5,6]

The fractional change in resistance $\Delta R/R$ (from parallel to anti-parallel alignment of the two identical FM layers) can be approximated by the Julliere formula[7]: $\Delta R/R = 2P_T^2/(1-P_T^2)$, where $P_T$ is the spin polarization of FM layers[8]. Importantly, $P_T$ is the *transport* spin polarizations, as it includes spin-dependent tunneling matrix elements and has to be distinguished from the density of states (DOS) spin polarization, $P_0 = [N\uparrow(E_F)-N\downarrow(E_F)]/[(N\uparrow(E_F)+N\downarrow(E_F)]$, where $N\uparrow\downarrow(E_F)$ is the DOS for the majority (minority) bands at the Fermi level. For $SrRuO_3$ band structure calculations[9,10] result in almost identical spin-up and spin-down DOS: $N_{\uparrow\downarrow}(E_F) \sim 23$ st/Ry, and, therefore $P_0 \approx 0$, even though SrRuO$_3$ has a relatively large magnetic moment.

Until recently, the only practical way to independently determine $P$ has been to use the tunneling technique pioneered by Tedrow and Meservey.[11] This technique is somewhat limited, however, by the need to grow a uniform tunnel barrier on top of one of the FM electrode. Soulen et al.,[12] and Upadhyay et al [13] have shown that transport spin polarization, $P$ can be determined from the current-voltage (*I-V*) characteristics of contacts formed between a superconductor and a ferromagnet. The character of *I-V* characteristics due to Andreev reflection at the interface[14] is quite different for spin-polarized and non-polarized parts of the current[15]. The PCAR technique, which emerged as a result of this work[12,13], has been used to measure the transport spin polarization of conventional magnetic systems, such as Ni-Fe alloys[16], as well as several oxide materials, such as CrO$_2$[12,17,18] and La$_{0.7}$Sr$_{0.3}$MnO$_3$[19].

In this Letter we present the measurements of the *transport* spin polarization of epitaxial thin films of SrRuO$_3$ using PCAR and compare them with the results of band structure calculations.

The films with the thickness 1200 Å were grown on miscut (2 degrees) (001) SrTiO$_3$ substrates by 90° off-axis sputtering.[20,5] Four-circle x-ray diffraction analysis indicated that these epitaxial SrRuO$_3$ thin films are single domains with the [110] direction normal to the substrate surface. The surfaces of the SrRuO$_3$ films consist of atomically smooth terraces with nearly periodic ledges along the miscut, resulting from coherent single domain growth.[21] The film surface is chemically very stable, which prevented any surface degradation and allowed us to obtain reproducible results in many consecutive spin polarization measurements.

In this experiment, we have used both as grown (strained) and lifted-off (freestanding, strain-free) films, similar to the ones described in Ref. 6. The in-plane lattice parameters of the as grown films are smaller than that of bulk SrRuO$_3$ (3.93Å), indicating that the film is subjected to a biaxial compressive strain in the plane of the film ($\varepsilon_{xx} = \varepsilon_{yy}$ =-0.64%). From the 2θ value of the normal scan the out-of-plane lattice parameter was found to be 3.96Å, which is larger than that of bulk materials, demonstrating a uniaxial tensile strain along [110] direction ($\varepsilon_{zz} = 0.50\%$) in the film. In contrast, both the in-plane and out-of-plane lattice parameters (~3.93Å) of the freestanding films are the same as that of the bulk material. This implies that the as grown SrRuO$_3$, which is initially subjected to an elastic strain, is fully relaxed after films are lifted-off. We found that the resistively measured Curie temperature, T$_C$ significantly increased from about 150 K for the strained film to about 160 K when the elastic strain was relaxed, which is consistent with our magnetization measurements.

Spin-polarization in the PCAR experiments can be represented by the expression:

$$P_n = \frac{I_\uparrow - I_\downarrow}{I_\uparrow + I_\downarrow} = \frac{N_\uparrow(E_F)v_{F\uparrow}^n - N_\downarrow(E_F)v_{F\downarrow}^n}{N_\uparrow(E_F)v_{F\uparrow}^n + N_\downarrow(E_F)v_{F\downarrow}^n} \quad (1),$$

where $v_{F\uparrow\downarrow}$ is the Fermi velocity of the majority (minority) spins, where n is either 1 or 2, depending upon whether conduction in the contact is in the ballistic (mean free path $L$ is much larger than the size of the contact $d$, $L>>d$) or diffusive ($L<<d$) regime, respectively[12,16,22]. Using the values of the resistivity of SrRuO$_3$ (~ 40 $\mu\Omega$·cm at 4 K) we can estimate the mean free path for minority (↓) and majority (↑) carriers, using the Ziman formula for $\sigma_\downarrow = \frac{1}{3}N_\downarrow(E_F)v_{F\downarrow}^2\tau_\downarrow$ and $\sigma_\uparrow = \frac{1}{3}N_\uparrow(E_F)v_{F\uparrow}^2\tau_\uparrow$ respectively and taking $\sigma = \sigma_\downarrow + \sigma_\uparrow$, from where $L \sim$ 80 Å and 45 Å for minority and majority carriers respectively. Using an interpolation formula[23] $R_n \approx \frac{4}{3\pi}\frac{\rho L}{d^2} + \frac{\rho}{2d}$ the contact size $d$ is estimated to be 50-150 Å, depending on the contact resistance. Thus both majority and minority carriers are in an intermediate regime of $L \sim d$ [24].

We studied four SrRuO$_3$ samples. Approximately 10 different contacts were measured for each sample. Soft Sn tips were used in to prevent any possible additional strain in the films due to the mechanical sample-tip interaction. Measurements of the *I-V* and differential conductance $G = (dI/dV)$ characteristics were made using a conventional four-terminal probe arrangement with a standard lock-in technique[12]. The values of *P* are extracted by fitting each individual data set with the recent theory[25] using a modified BTK model[26], generalized to include a spin-polarized metal.[25] The fitting least square routine includes corrections for "spreading" film resistance $R_s$[27] and then varies only two parameters, *P* and the strength of the BTK delta-function, *Z*.[26] The temperature is taken to be the measured physical temperature of the film (typically ~1.6 K), while the superconducting gap *Δ* is taken from the BCS temperature dependence for Sn. The ballistic model can easily fit most of the contacts. This is consistent with our previous work on La$_{0.7}$Sr$_{0.3}$MnO$_3$[19], in which we have found that the intermediate regime is closer to the ballistic than to the diffusive limit.

Figures 1 (a-b) show examples of normalized conductance data and the fits for one of the as grown and freestanding samples respectively. Table I gives a summary of the results for all four samples. The average value of *P* is 52.5% with no obvious difference between the two types of films. However, a standard deviation, *ΔP* for each film and the spread in the values of *P* among the films, seem to be larger for the as grown than for the freestanding films. This may indicate that the results for the freestanding films are more representative of the bulk material's *P*, while the rather large variation in the values of *P* for the as grown films might be due to the film anisotropy and the effects of local strain.

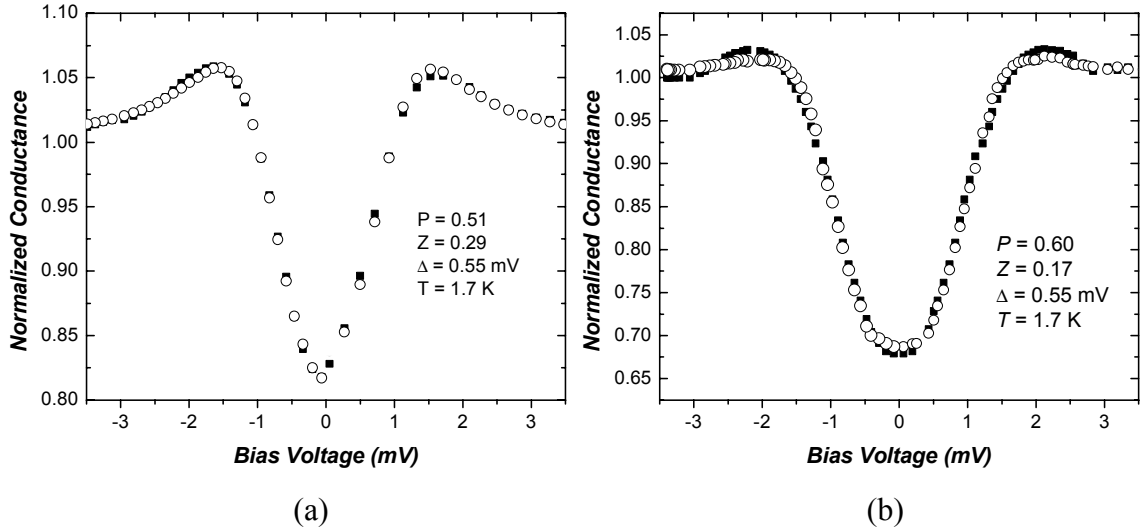

(a)                    (b)

Figure 1a-b. Examples of conductance curve (normalized for $V \gg \Delta/e$) of at 1.7 K, fitted with the least square fit routine based on the theory (ballistic limit) of Ref. 25. The solid squares are the experimental data, the open circles - the fits; a) sample #2 (as grown) contact 12 ($P$ = 51%; $Z$ ~0.3). b) Sample # 4 contact 13 ($P$ = 60%, $Z$ ~ 0.17).

Importantly, the chemical stability of the films resulted in (generally) very low interface barrier *Z*, 0<*Z*<0.3. Thus our measurements have been practically unaffected by any possible dependence of the values of *P* on *Z*, which have been reported in some systems. [17] We would like to emphasize, that, even though this dependence is, in principle, possible due to spin-dependent scattering or strong spin-orbit coupling, one has to be very careful not to confuse a finite *Z* in the ballistic limit with the *Z* = 0 case in the diffusive limit. To illustrate this point we present an example (in Figures 2a –2b) of one out of the two contacts we have measured, which could have been described by a higher *Z*. While it is practically impossible to fit this data by the ballistic theory *unless* one starts to vary *Δ* (which in this case has to be 30% higher than the BCS value), we can easily fit the same data with the diffusive theory, with the correct *Δ* and *Z* = 0.

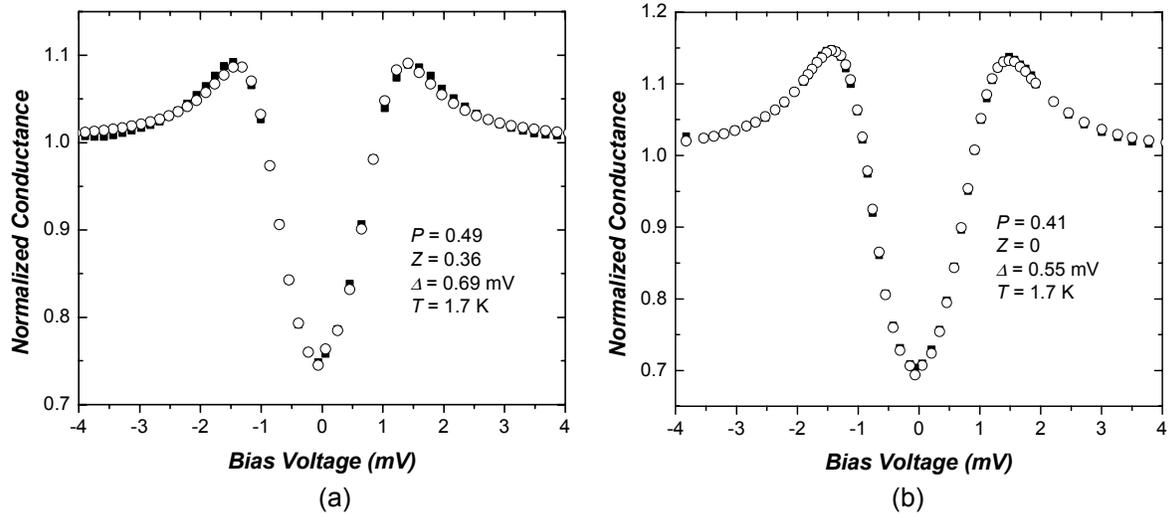

Fig. 2 a) *Ballistic* fit with $Z = 0.36$ and inflated gap ($\Delta = 0.69$ mV); b) *Diffusive* fit of the same data (#2 contact 19) with the correct gap ($\Delta = 0.55$ mV) and $Z = 0$.

To compare the experimental data with the electronic band structure results the density functional calculations, using the general potential linearized augmented planewave method as previously described[28, 9], were done for SrRuO$_3$ in the orthorhombic *Pbnm* structure[29]. This structure has a primitive unit cell consisting of four SrRuO$_3$ formula units arising from a $\sqrt{2} \times 2 \times \sqrt{2}$ superstructure of the simple perovskite cell. This superstructure consists of rotations of the oxygen octahedra accompanied by small lattice strains. In particular, the lattice parameters of that cell are 5.5304 Å, 7.8446 Å, and 5.5670 Å, so the *c* axis is 0.7 % longer than the *a* axis[30]. In order to compare with experiment, we calculated the Fermi surface averages $<v>$ and $<v^2>$ along the *a* and *c* directions (see Table II). These averages were found by a very fine interpolation of the energy bands around the Fermi energy, based on 567 first principles *k*-points in the irreducible wedge of the Brillouin zone. The average measured values of $P \sim 52.5\%$ are in good agreement with the predicted spin polarization for a diffusive contact.

The fact that we used mainly ballistic theory to fit the data and yet our results seem to yield much better agreement with the theoretical calculations in the diffusive limit may look somewhat surprising. We note in this regard that the formulas used to describe conductance in the spin-polarized case[25] consist of different pre-factors and functional dependencies $\langle Nv \rangle$ and $f_b(V)$ in the ballistic case and $\langle Nv^2 \rangle$ and $f_d(V)$ in the diffusive case respectively. It is quite possible, that a transition from $\langle Nv \rangle$ to $\langle Nv^2 \rangle$ happens at different values of $L/d$ than a transition from $f_b(V)$ to $f_d(V)$, which would explain this result.

Using tunneling spectroscopy D.C. Worledge and T.H. Geballe[4] found that the spin polarization of $SrRuO_3$ is ~ -10%[31], significantly smaller in absolute value than the spin polarization measured in this work. The authors of Ref. 4 discussed several possible explanations for this discrepancy between their measurements and our results. In our opinion the major factor is the difference in the barrier decay length for different surface states[32]. This only reinforces our belief that comparison of $P$ measured by different techniques should be correlated with specific experiments.

In conclusion, we have done PCAR measurements of the spin polarization of the itinerant ferromagnet $SrRuO_3$. The spin polarization results for freestanding films are more uniform than for as grown films and are likely to reflect the properties of the bulk material. Our average experimental value of $P$ (~52.5%), is in good agreement with *LAPW* calculations in the diffusive limit, thus demonstrating that the transport spin polarization in this material is primarily due to the difference in the Fermi velocities of the majority and minority spin carriers and can be dramatically different from the DOS spin polarization.[33]

We thank I.I. Mazin and for numerous discussions and for developing the fitting routine, and J. Claassen for technical assistance.

| Sample | P (%) | ΔP (%) |
|---|---|---|
| #1 (as grown) | 43 | 6.0 |
| #2 (as grown) | 54 | 4.0 |
| #3 (freestanding) | 55 | 2.0 |
| #4 (freestanding) | 58 | 3.5 |

Table I. Summary of the spin-polarization results for the four SrRuO$_3$ samples studied, *P* is the average spin-polarization, and ΔP is the one-standard deviation uncertainty in *P*.

| | $\langle |v_a| \rangle$ | $\langle |v_b| \rangle$ | $\langle |v_c| \rangle$ |
|---|---|---|---|
| Majority | 0.60x10$^7$ cm/s | 0.69x10$^7$ cm/s | 0.62x10$^7$ cm/s |
| Minority | 1.27x10$^7$ cm/s | 1.07x10$^7$ cm/s | 1.25x10$^7$ cm/s |
| $P_{Nv} = P_v$ | -0.34 | -0.22 | -0.36 |

| | $\sqrt{\langle v_a^2 \rangle}$ | $\sqrt{\langle v_b^2 \rangle}$ | $\sqrt{\langle v_c^2 \rangle}$ |
|---|---|---|---|
| Majority | 0.83x10$^7$ cm/s | 0.86x10$^7$ cm/s | 0.86x10$^7$ cm/s |
| Minority | 1.59x10$^7$ cm/s | 1.57x10$^7$ cm/s | 1.57x10$^7$ cm/s |
| $P_{Nv}^2 = P_v^2$ | -0.57 | -0.54 | -0.54 |

Table II. Calculated majority and minority Fermi velocities and the spin polarization $P_{Nv}$ (top) and $P_{Nv}^2$ (bottom) with respect to the lattice directions *a,b,c* (see text). The densities of states are assumed to be the same for majority and minority electrons. Note the rather large anisotropy for $P_{Nv} = P_v$ along *b* and *a/ c* axis. The minus sign of the spin polarization reflect the fact that the minority spin-current is larger in both ballistic and diffusive cases. This cannot be directly confirmed by the PCAR technique, which can only measure the absolute value of the transport spin polarization but is consistent with the results of Ref. 4.